\documentclass[3p,twocolumn,sort&compress]{elsarticle}
\usepackage{graphicx} % needed for figures
\usepackage{dcolumn} % needed for some tables
\usepackage{bm} % for math
\usepackage{amssymb} % for math
\usepackage{amsmath} % for align environment

\usepackage{xcolor}
\usepackage{subfig}

\definecolor{darkgreen}{RGB}{50,150,0}

\usepackage[most]{tcolorbox}
\newtcolorbox{highlighted}{colback=cyan,coltext=white,breakable}
\usepackage{lipsum}

\begin{document}

\title{On the Cosmological Implications of the String Swampland}
\author[hvd]{Prateek Agrawal}
\author[hvd]{Georges Obied}
\author[pri,ccp]{Paul J. Steinhardt}
\author[hvd]{Cumrun Vafa}
\address[hvd]{Jefferson Physical Laboratory, Harvard University
17 Oxford Street, Cambridge, MA 02138, USA}
\address[pri]{Department of Physics \& Princeton Center for Theoretical Science, Princeton University,
Princeton, NJ 08544, USA}
\address[ccp]{Center for Cosmology and Particle Physics, Department of Physics, New York University,
New York, NY, 10003, USA}
\date{\today}

\begin{abstract}
  We study constraints imposed by two  proposed string
  Swampland criteria on cosmology.  These criteria involve an upper
  bound on the range traversed by scalar fields as well as a 
  lower bound on
  $|\nabla_{\phi} V|/V$ when $V >0$.   We find that inflationary
  models are generically in tension with these two criteria. Applying
  these same criteria to dark energy
  in the present epoch, we find that specific quintessence models can
  satisfy these bounds  and, at the same time, satisfy
  current observational constraints.  Assuming the two Swampland
  criteria are valid, we argue that the universe will undergo a phase
  transition 
  within a few Hubble times.
  These criteria
    sharpen the motivation for future
    measurements of the tensor-to-scalar ratio $r$ and the dark energy 
    equation of state $w$, and for tests of the equivalence principle
    for dark matter.
\end{abstract}

%\pacs{}
\maketitle

\section{Introduction}

The landscape of
string theory gives a vast range of choices for
how our universe may fit in a consistent quantum theory of gravity.
However, it
is believed that this is surrounded by an even bigger Swampland,
i.e., a set of consistent looking effective quantum field theories
coupled to gravity, which are inconsistent
with a quantum theory of gravity~\cite{Vafa:2005ui}. For a recent
review of the subject and references see~\cite{Brennan:2017rbf}.  The
aim of this paper is to investigate the cosmological implications of two
of the proposed Swampland criteria.

The two Swampland criteria whose consequences we will study are:

\noindent
{\sc Criterion 1}:  {\it  The range traversed by scalar fields in field space is
  bounded by $\Delta \sim {\cal{O}}(1)$ in reduced Planck units}
\cite{Ooguri:2006in}.  More precisely, consider a theory of quantum
gravity coupled to a number of scalars $\phi^i$ in which
the effective Lagrangian, valid up to a cutoff scale $\Lambda$, takes the form
\begin{align}
  \label{lagrange}
  \mathcal{L}
  &=
  \sqrt{|g|} 
  \left[\frac{1}{2} R
    -\frac{1}{2}
    g^{\mu\nu}\partial_{\mu} \phi^i \partial_{\nu} \phi^j
    G_{ij}(\phi)
    \right.\nonumber\\&\qquad\qquad\qquad\qquad\qquad\qquad\left.
    -V(\phi)
    +\ldots
    \frac{}{}
  \right],
\end{align}
where we use reduced Planck units throughout.  Note in particular that
we go to the Einstein frame and use $G_{ij}(\phi)$ in this frame to
define a metric which we use to measure distances in the field space
$\phi^i$.
Then it is believed that there
is a finite radius in field space where the effective Lagrangian above
is valid.  In particular if we go a large  distance $D$ in field
space, a tower of light modes appear with mass scale
\begin{align}
 m\sim M_{pl}\exp(-\alpha D)
\end{align}
which invalidates the above effective Lagrangian.  Here $\alpha \sim
{\cal O}(1)$.  This means that any effective Lagrangian has a proper
field range\footnote{Note that the proper field range is measured
  along the
minimum loci of the potential for a given effective cutoff scale.} for
$|\Delta \phi|<\Delta$, where the expectation is that $\Delta \sim
{\cal O}(1)$.  There is by now a lot of evidence for this conjecture.
See in
particular~\cite{Grimm:2018ohb,Heidenreich:2018kpg,Blumenhagen:2018hsh}
for a recent discussion and extensions
of this conjecture.

{\sc Criterion 2}:  {\it There is a lower bound $|\nabla_{\phi} V|/V
>c \sim {\cal O}(1)$ in reduced Planck units  in any consistent theory of
gravity when $V>0$.} The second Swampland criterion, which was
recently conjectured in \cite{Obied:2018fos}, is motivated by the observation
 that it appears to be difficult to construct any reliable dS vacua and
by experience with string constructions of scalar potentials.

We will be applying these two criteria to periods of possible cosmic
acceleration.  In particular we revisit
early universe inflationary models in view of these constraints as
well as study
their implications for the present epoch (dark energy), and our
immediate future.  Since the values of $c$ and $\Delta$ are not
precisely known,
the best
we can do is to
formulate constraints in terms of these unknown constants.  

We find that inflationary models are generically in tension with these
two criteria depending on how strictly we interpret these constraints
in terms of the proximity of $\{c, \Delta\}$ to $1$.  For example,
among inflationary models that are not ruled out by current
observations, plateau models require $c\lesssim 0.02$ and $\Delta \gtrsim 5$.

As for the present universe, the second Swampland criterion is clearly
in conflict with $\Lambda$CDM cosmology  because a positive
cosmological constant violates the bound $|\nabla_{\phi} V|/V >c>0$.  
However, quintessence models of dark energy~\cite{Caldwell:1997ii} can
be made consistent
with the  two criteria.
Aside from the inflationary constraints, considering only cosmological
observations of the recent universe, 
we derive model-independent constraints on the
values of $\{c,\Delta\}$, 
$c<0.6$ and $c <3.5\ \Delta$. These values can be
realized in concrete quintessence models.
Moreover we find a lower bound on the deviation of today's value of
$w$ from $-1$
given by $(1+w)\gtrsim 0.15\ c^2$, where $w=p/\rho$ for the dark
energy component of the universe.

Extrapolating these models to the future, we find that in a time of
order $t_{end}{ \lesssim [{\frac{3\Delta}{2 c\Omega_\phi^0}}}] \ H_0^{-1}$
the universe must enter a new phase. 
Here $H_0$ is the current value of the Hubble parameter and
$\Omega_\phi^0=0.7$ is the current density fraction of dark energy.
So $t_{end}$ could be viewed as
``the end of the universe as we know it'' and the beginning of a new
epoch.  The new epoch may entail the appearance and production of a
tower of light states and/or the transition from accelerated expansion
to contraction.

The organization of this paper is as follows:  We first discuss
constraints on early universe
inflationary models
and then discuss how the recent and present cosmology fits with the
above criteria.  Finally
we discuss the future of our universe in view of these criteria.

\noindent
\section{Past}
Observational constraints on
inflation~\cite{Guth:1980zm,Linde:1981mu,Albrecht:1982wi}, the
 hypothetical period of cosmic acceleration in the very
early universe,
are in tension with both Swampland criteria. The tension with
Criterion 1 ($\Delta \lesssim 1$) has been noted
previously~\cite{Banks:2003sx,ArkaniHamed:2006dz,
Silverstein:2008sg,Heidenreich:2015wga},
but the tension with Criterion 2 ($|\nabla_{\phi}V|/V \geq c \sim {\cal
O}(1)$) has not been studied before.

Let us first briefly review some of the parameters of inflationary models relevant for these criteria.
Consider single-field slow-roll inflation based on an action of the form shown in Eq.~(\ref{lagrange}). In the slow-roll limit, the equation of state is
\begin{equation}
\epsilon \equiv \frac{3}{2}(1+w) \equiv \frac{3}{2}\left(1+ \frac{p}{\rho}\right) \approx \frac{1}{2}\left(\frac{|\nabla_{\phi}V|}{V}\right)^2,
\end{equation}
where $p=\frac{1}{2} \dot{\phi}^2 -V$ and $\rho=\frac{1}{2} \dot{\phi}^2 +V$ are the homogeneous pressure and energy density, respectively.

The relation between $\epsilon$ and $N_e$, the number of $e$-folds remaining before the end of inflation, is
\begin{equation} \label{eps}
\epsilon \sim \frac{1}{N_e^{k}}.
\end{equation}
The exponent $k$ is equal to 1 for inflationary potentials in
which $V(\phi)$ scales roughly as an exponential or  power-law to
leading order in $\phi$ during inflation, which includes models with
the fewest parameters and least fine-tuning; and equal to 2 for a
special subclass of more fine-tuned ``plateau models" in which
$V(\phi)$ is nearly constant during inflation and ends inflation with
a sharp cliff-like drop to a minimum. During the last $N_e$ $e$-folds,
the range of $\phi$ is roughly~\cite{Garcia-Bellido:2014wfa}
\begin{equation} \label{Delta}
\Delta \phi \sim N_e \sqrt{2 \epsilon} \sim \sqrt{2} N_e^{1-k/2}.
\end{equation}

We begin by considering constraints on $V(\phi)$ during the last $N_e \approx 60$ $e$-folds, the period probed directly by measurements of the cosmic microwave background.
The exponential- and power-law-like inflationary models are ruled out by recent observational
limits on B-mode polarization that constrain the tensor-to-scalar
fluctuation amplitude ratio $r \approx 16 \epsilon < 0.07$ or
$\epsilon < 0.0044$~\cite{Array:2015xqh}. This, combined with
measurements of the spectral tilt $n_s$ of the scalar density
fluctuations, is incompatible with these inflationary models
(which all have $\epsilon \gtrsim 0.01$).  However, current
constraints allow the more fine-tuned plateau models
(with $\epsilon < 0.0005$)~\cite{Ijjas:2013vea}.

We now turn to evaluating the two Swampland criteria for the past
cosmic acceleration (inflation) which turns out to be difficult
to satisfy for several reasons:
\begin{enumerate}
\item the period of acceleration must be maintained for many $e$-folds of expansion;
\item there are many different observational constraints to be
  simultaneously satisfied (on tilt, tensor-to-scalar ratio,
  non-gaussianity, and
  isocurvature perturbations);
\item the empirical constraints are quantitatively tight.
\end{enumerate}

\noindent
{\sc Criterion 1:} Based on Eq.~(\ref{Delta}), we see that the range
of $\phi$ spanned during the last $N_e =60$ $e$-folds is ${\cal O}(1)$ or
greater. Plateau models have the least tension with Criterion 1, but,
even in these cases, when factors of order unity are fully included,
the range is $\Delta \ge 5$ in reduced Planck mass units. While the
tension may be viewed as modest, we note that the range can be much
larger if there are more than the minimal 60 e-folds of inflation.

\noindent
{\sc Criterion 2:} The current B-mode constraint $\epsilon < 0.0044$
corresponds to $|\nabla_{\phi} V|/V < 0.09$, in tension with the
second Swampland criterion $|\nabla_{\phi} V|/V > c \sim {\cal{O}}(1)$. 
Near-future measurements will be precise enough to detect values of $r$ at
the level of 0.01; failure to detect would require $|\nabla_{\phi}
V|/V \lesssim 0.035$. The plateau models, favored by some
cosmologists
as the simplest remaining that fit current observations, require
$|\nabla_{\phi} V|/V \lesssim 0.02$ during the last 60 $e$-folds,
which is in greater tension
with the second Swampland criterion.

Hence, we see generically current observational constraints on
inflation are already in modest tension with the first Swampland
criterion and more so with the second Swampland
criterion especially in the context of the plateau models, which are
observationally favored.  Near-future experiments can further
exacerbate the tension
if they place yet tighter bounds on $r$.

Note that we have only considered thus far the tension with Criterion
2 during the last 60 $e$-folds. In practice, nearly all inflationary
models in the literature include extrema or plateaus or power-law
behavior in which $|\nabla_{\phi} V|/V \rightarrow 0$
at one or more values of $\phi$. These are forbidden by Swampland
Criterion 2.

Variants of single-field slow-roll inflation do not provide any
apparent relief and/or run into other observational constraints. DBI
inflationary models replace the kinetic energy density of the inflaton
with a Born-Infeld action~\cite{Silverstein:2003hf}. In this case, the Swampland criteria apply by first taking the limit of small $(\partial_{\mu} \phi)^2$ and normalizing fields so that the kinetic energy density is canonical. If $(\partial_{\mu} \phi)^2 \ll 1 $ throughout inflation, the constraints above apply directly. In cases where
$(\partial_{\mu} \phi)^2 $ becomes order unity during inflation, the model runs into  constraints on non-gaussianity.

For Higgs inflation~\cite{Bezrukov:2007ep}, $R^2$
(Starobinsky) inflation~\cite{Starobinsky:1980te},
pole-inflation~\cite{Broy:2015qna} and $\alpha$-attractor
models~\cite{Kallosh:2013hoa}, evaluating the Swampland criteria
requires first redefining the metric and scalar fields  such that the
action is recast in the form of Einstein gravity plus a canonical
kinetic energy density for the scalar field. In this form, they all
correspond to plateau models which, as shown above, are in modest
tension with Criterion 1 and in significant tension with Criterion 2
at $N_e =60$ (and in even greater tension for larger $N_e$ because
$|\nabla_\phi V/V| \rightarrow 0$).

Axion monodromy models~\cite{Silverstein:2008sg},
N-flation~\cite{Dimopoulos:2005ac} and other multifield
models were introduced to ensure that no field traverses a linear
field distance from the origin greater than unity. However, as noted
in the introduction, Swampland Criterion 1 is based on the total path
length along the slow roll trajectory (more precisely, along a
gradient flow trajectory) in the field space. The
strategies are not sufficient to satisfy Criterion 1 if the total path
length exceeds order unity, which is the situation in these cases.

The more serious tension, though, is nearly always with Swampland Criterion 2.
Almost all inflationary constructions include extrema or plateaus in
which $|\nabla_{\phi} V|/V \rightarrow 0$ at one or more points in
field space. It remains a challenge to find examples that satisfy
observations and also satisfy $|\nabla_{\phi} V|/V > c \sim {\cal
O}(1)$. If one cannot be found, there are only a few options. Either
the Swampland criteria are wrong, which can be proven by a full
construction of counterexamples; or inflation cosmology is wrong and
some other mechanism accounts for the smoothness, flatness and density
perturbation spectrum of the observable universe\footnote{Alternative
ideas  include string gas cosmology and bouncing cosmologies.}; or
perhaps both are deficient and theoretical and observational progress
will point to new possibilities.

\medskip
\noindent
\section{Present}
Current data shows that the universe is dominated by dark energy.  
Criterion 2 already implies that this cannot be the result of a positive
cosmological constant or being at the minimum of a potential with
positive energy density, and so we must be dealing with a scalar field
potential that is rolling, i.e. a quintessence model.  
 Furthermore, if there are
generic string compactifications that predict a particular lower bound for
$|\nabla_\phi V/V|$, this implies that the slope of the potential is
naturally small when the dark energy is small, perhaps putting
quintessence on a firmer theoretical footing, even without assuming
the validity of the second Swampland criterion.

However in
string theory scalar fields typically determine coupling constants and
at first this may appear to be in tension with the fact that, for
example, the change in the fine structure constant  is $\lesssim
10^{-6}$ out to redshift $z= {\cal O}(1)$ \cite{Martins:2017yxk}.  But
as pointed out in \cite{Brennan:2017rbf} this simply means that the
scalar fields should couple to some other fields other than the
visible matter.  In other words, this anticipates the existence of the
dark matter sector to which they should be more strongly coupled.  In
string theory such a scenario would be realized by models where the
standard model arises from a localized region of
internal geometry (such
as in F-theory model building), whereas dark matter could arise from
some other regions.  In this context the quintessence field would
correspond to the volume of the other region where dark matter
originates and thus may control the couplings in the dark matter sector.

Astrophysical observations that constrain the ratio of the dark
energy density to the critical density ($\Omega_{\phi}(z)$) and
equation of state ($w(z)$) of dark energy as a function of redshift
($z$) can be used to test Criterion 2.  One of the features of
quintessence models is that not only is the value of $V$ small (of
the order of $10^{-120}$ in reduced Planck units) but its slope $V'$ should
also be small and again of order $10^{-120}$ (or less) in reduced Planck
units.  Intriguingly, Criterion 2 gives at its boundary a value of
$V'$ of the same order as $V$.  This  relationship means that current
experiments already impose bounds on the value of $c$ in Criterion 2
and future experiments have the possibility of significantly
tightening those bounds.

\begin{figure*}[ht]
  \centering 
  \subfloat[a][\label{fig:a}]
  {\includegraphics[width=0.45\textwidth]{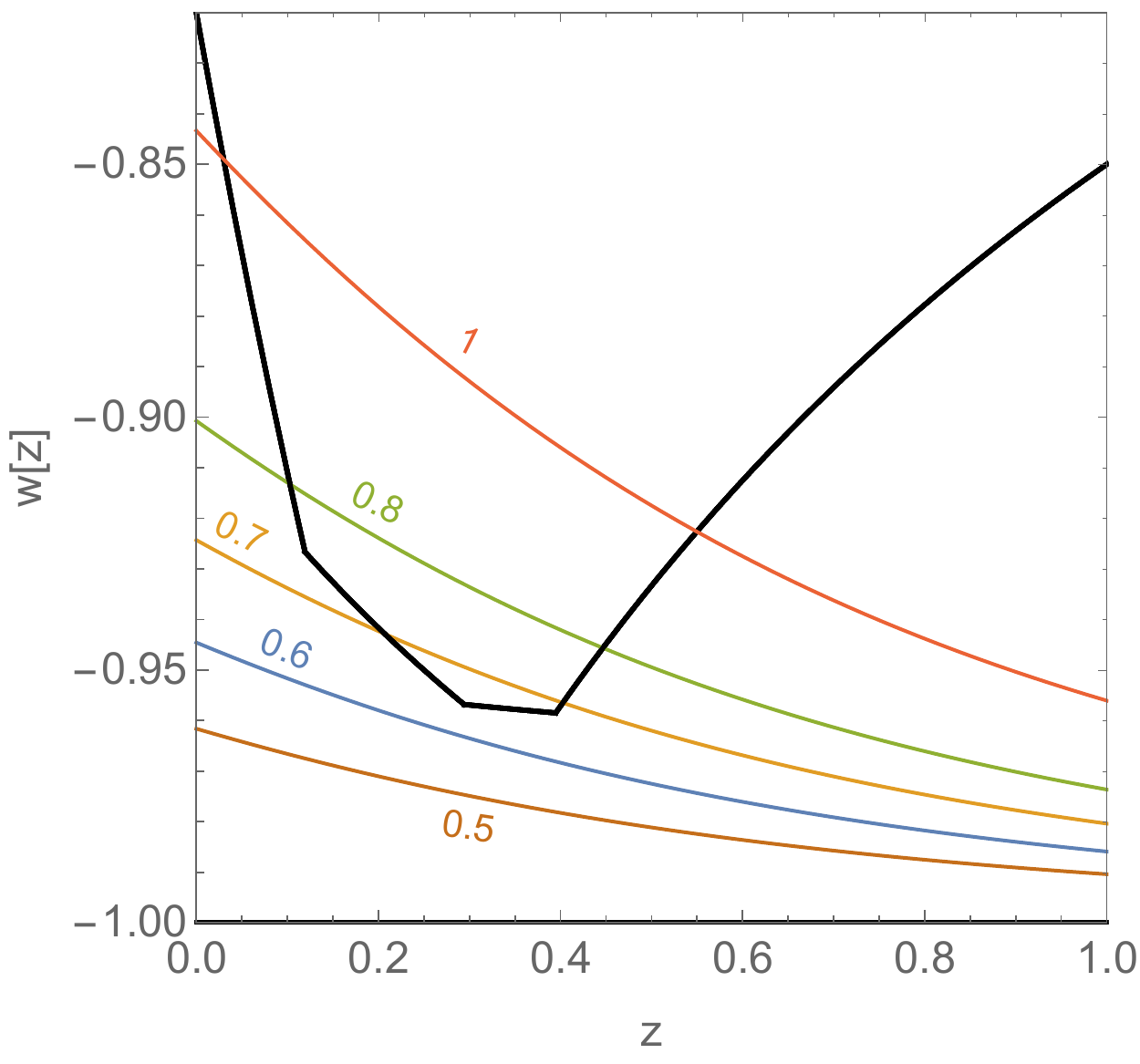}} \quad
  \subfloat[b][\label{fig:b}]
  {\includegraphics[width=0.435\textwidth]{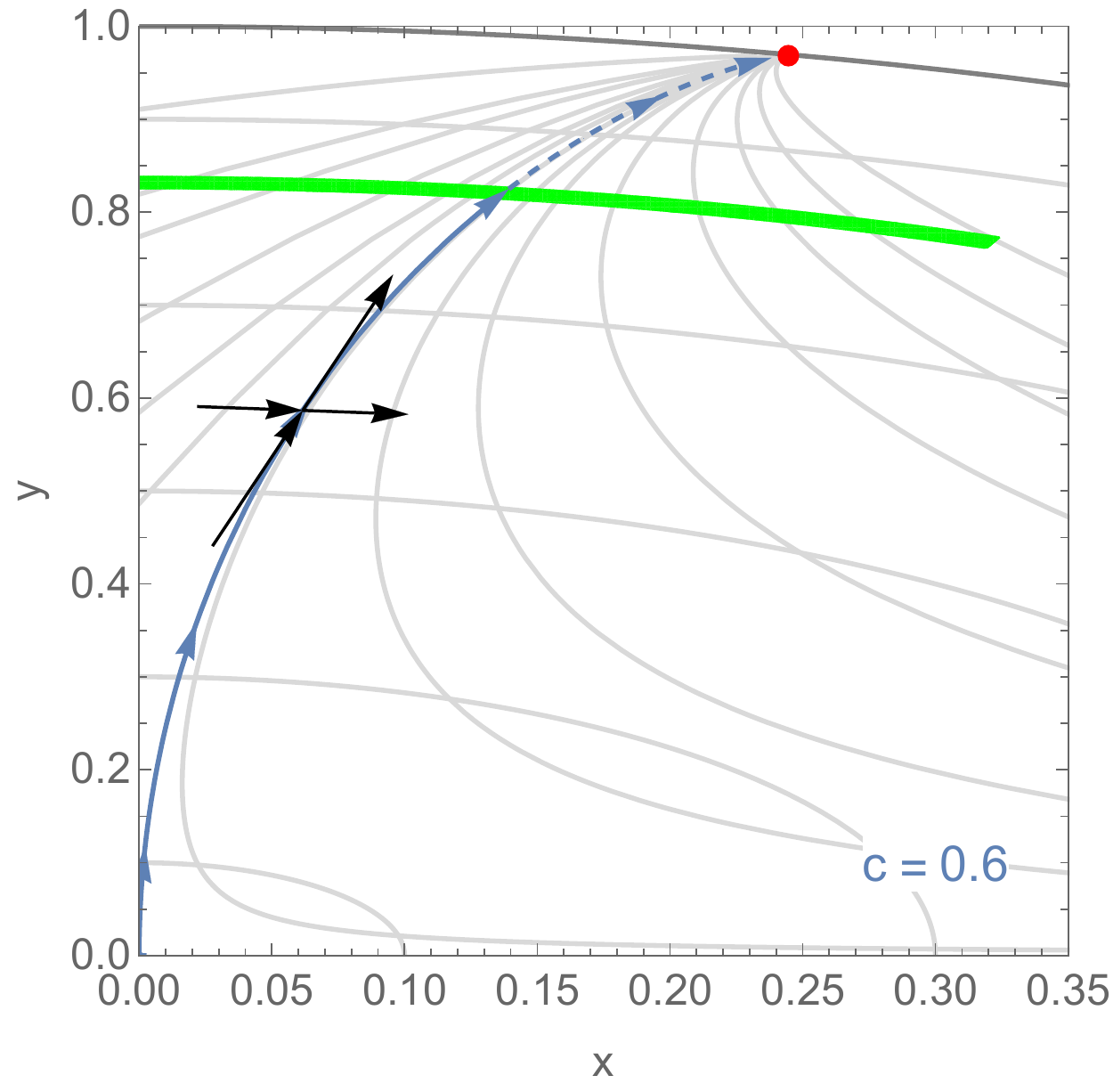}} 
  \caption{
    (a) The black curve shows the current observational
    $2\sigma$ bound on $w(z)$ for $0<z<1$ based on SNeIa, CMB and BAO
    data \cite{Scolnic:2017caz}.    This is compared with the
    predicted $w(z)$ for exponential quintessence potentials with
    different values of constant $\lambda$ under the constraint that
    $\Omega_{\phi}(z=0)= 0.7$ and assuming initial conditions
    $x=y\approx 0$. From this we observe that the upper bound on
    $\lambda$ is $\sim 0.6$ (blue curve).  
    (b) The blue
    curve shows the trajectory in the $(x,y)$ plane corresponding to
    constant $\lambda=0.6$, the upper bound allowed in Fig.~1(a),
    assuming initial conditions $(x,y)=(0,0)$.  The current $(x,y)$ is
    where the blue curve meets the green; the dashed blue curve
    illustrates its future asymptotic behavior.  
    Trajectories to the right of the blue curve have a larger
    $w(z)$ at $0<z<1$ and, hence, violate the
    observational constraints in Fig.~1(a).  As explained in the
    text, trajectories to the left of the blue curve extrapolate back
    in time, hit the $y$-axis at some finite $y$ and then continue on
    to $(x,y)=(-1,0)$ or $\Omega_{\phi} \rightarrow 1$.  These
    trajectories disrupt matter domination and, hence, large-scale
    structure formation.  Hence, the bound for constant $\lambda$,
    $c<0.6$ in Fig.~1(a), is also the bound for general
  $\lambda(\phi)>c$.   }
  \label{fig:sol}
\end{figure*}

We consider here current constraints from supernovae (SNeIa), cosmic
microwave background (CMB) and baryon acoustic oscillation (BAO)
measurements given in Ref.~\cite{Scolnic:2017caz}.  These require
that:
\begin{itemize}
\item $1+w(z) \ll 2/3$ for $z<1$ (see Ref.~\cite{Scolnic:2017caz} and Fig.~\ref{fig:sol}); 
\item $\Omega_{\phi}(z=0) 
\equiv \Omega_{\phi}^0 
  \approx 0.7$ \cite{Ade:2015xua}; and
\item $\Omega_{\phi}(z>1) \ll 1$ in order to avoid suppression of large-scale structure formation.
\end{itemize}
While a model-by-model comparison to data would give the most precise bounds, the approach employed here is sufficiently accurate for our purposes of obtaining bounds on the parameters  $\Delta$ and $c$ that appear in the two Swampland criteria.

For a canonically normalized field $\phi$, the field trajectory can be conveniently parameterized by the dynamical variables 
\begin{align}
  x &= \frac{\dot\phi}{\sqrt{6} H}
  \\
  y
  &=
  \frac{\sqrt{V(\phi)}}{\sqrt{3} H }
\end{align}
where $-1<x<1,0<y<1$.
In terms of these variables,
\begin{align}
  \Omega_\phi
  &=
  \frac{\frac{1}{2}\dot{\phi}^2+ V(\phi)}{3 H^2}
  = x^2 + y^2
  \\
  1+w &=
  \frac{2x^2}{x^2+y^2},
  \label{eq:w}
\end{align}

The equations of motion in terms of $x$ and $y$ are (see
\cite{Tsujikawa:2013fta} for a recent review),
\begin{align}
  \frac{dx}{dN}
  &=
  \frac{\sqrt{6}}{2} \lambda y^2
  -3x
  +\frac32 x \left[(1-w_m)x^2 
  \right.\nonumber\\&\qquad\qquad\qquad\qquad
  +  \left.(1+w_m) (1-y^2)\right]
  \\
  \frac{dy}{dN}
  &=
  -\frac{\sqrt{6}}{2} \lambda x y
  +\frac32 y \left[(1-w_m)x^2 
  \right.\nonumber\\&\qquad\qquad\qquad\qquad
  + \left.  (1+w_m) (1-y^2)\right]
  \label{eq:quintxy}
\end{align}
where $w_m$ is the equation of state of the other components of the
universe. Since we focus on the matter-dominated and  dark energy-dominated epochs,
 $w_m \simeq 0$.
Here $\lambda(\phi) \equiv |\nabla_{\phi}V|/V$. By Swampland
Criterion 2, $\lambda(\phi) \ge c \sim {\cal{O}}(1)$.  As we shall see below,
the data puts an upper bound on $c$.  To find 
this upper bound we proceed in two steps. 

First, we consider the special case of  exponential potentials with
constant $\lambda$:
\begin{align} V(\phi) &=  V_0 e^{\lambda \phi } \label{eq:constlambda}
\end{align}
The predictions of $w(z)$ for a given $\lambda$ depend in general on
the initial conditions. These are fixed by the
requirement that $\Omega_{\phi}(z)$ become negligible at $z
>1$, as needed for large-scale structure
formation. Therefore, in the far past we begin close to the repulsive
fixed point $(x,y)=(0,0)$, and start rolling towards
the fixed point at $(x,y) =
({\lambda}/{\sqrt{6}}, \sqrt{1-\lambda^2/6})$, such that
$\Omega_\phi^0 = 0.7$. In 
figure~\ref{fig:sol}\subref{fig:a}, we plot the $w(z)$ predictions
from these trajectories for a range of values of $\lambda$.
We compare these predictions with the current $2 \sigma$
upper bounds on $w(z)$ for $0<z<1$  (black curve\footnote{
    The black curve is determined from Fig.~21 in
    Ref.~\cite{Scolnic:2017caz} by finding the
    values of  $(w_0,w_a)$ all along the 2$\sigma$ contour; plotting
    all $w(z)$ of the form $w(z)=w_0 + w_a  z/(1+z)$; and finding
    the upper
    convex hull.
})~\cite{Scolnic:2017caz}.  The comparison shows that the
upper bound on $\lambda$ is  $0.6$, somewhat less than
unity.

Second, a universal upper bound on $c$ can be derived for general
$\lambda(\phi)$. We claim and will shortly prove
that the constant
$\lambda$ case with $\lambda(\phi) = c$ is the \emph{least}
constrained trajectory. From above, such a trajectory is ruled
out if $c$ is bigger than $0.6$. It follows that every possible
$\lambda(\phi)$ is ruled out if $c$ is bigger than $0.6$, leading to
the bound $c\lesssim0.6$.

We now provide the argument why the $\lambda(\phi) = c$ trajectory is
the limiting case.
Figure~\ref{fig:sol}\subref{fig:b} shows in blue the
trajectory for the
case $\lambda=c=0.6$ which 
connects the 
fixed point at $(x,y) = ({c}/{\sqrt{6}}, \sqrt{1-c^2/6})$ 
to  the repulsive fixed
point at $(x,y)= (0,0)$.  
From figure~\ref{fig:sol}\subref{fig:a}, this trajectory fits  observational
constraints for $0<z<1$ and $z>1$.  
Where the blue curve intersects the upper
black line in the future is the stable fixed point; if the universe
began at the repulsive fixed point in the past, the current position
along the trajectory is where the blue curve meets the green one.

Note that trajectories are bounded by the condition
\begin{align}
  -\frac{x}{y} 
  <
  \frac{dy}{dx}
  <
  -\frac{x}{y}
  \left(
  1
  -\frac{(1-x^2-y^2)(y^2-x^2)}
  {\sqrt{\frac23} c x y^2 - x^2 (1-x^2+y^2)}
  \right)
  \label{eq:slope}
\end{align}
where we use the fact that the slope $dy/dx$ for each trajectory at
each point is a monotonic function of $\lambda$.  
Starting from any point in the $x-y$ plot, we can use this condition
to bound any trajectory that passes through that point if
$\lambda(\phi)>c$.  Namely, draw trajectories through the point with
$\lambda(\phi)=c$ and $\lambda(\phi) \rightarrow \infty$; these form a
cone through which any other trajectory for general $\lambda(\phi)$
must pass. This is illustrated, for example,  by the black lines with
arrows in Fig.\ref{fig:sol}\subref{fig:b} and the grid of gray lines.  
Since the blue curve is along one of the edges of this cone, it follows that
any trajectory that passes through a point on the
right side of the blue curve will stay to the right and cannot cross
to its left in the future. Points to the right of the blue curve
correspond to larger values of $w(z)$ (equation~\eqref{eq:w}), and
hence all trajectories that are in this region are more constrained by
data as shown in figure~\ref{fig:sol}\subref{fig:a}.

Similarly, starting from any point to the left of the blue curve and 
extrapolating to the past, the
trajectory must remain to the left of the blue curve.  
These trajectories will intersect the
$y$-axis at some finite value of $y$ at some finite earlier time.  On the
$y$-axis, the kinetic energy density is zero, corresponding to 
a ``turning point''. On this trajectory,  the
field is initially rolling  uphill ($x<0$);  it stops at the turning
point ($x=0$, where the trajectory hits the $y$-axis); and then rolls down
hill ($x>0$).   The kinetic energy density increases  $\propto 1/a^6$
compared to matter $\propto 1/a^3$ or radiation $\propto 1/a^4$ as $a
\rightarrow 0$ going back in time.  Consequently, the kinetic energy
density rapidly grows to dominate over all forms of energy the further
back one extrapolates. The result is that any trajectory passing
through a point  on the left of the blue curve in
Fig.~\ref{fig:sol}\subref{fig:b}
traces back to $(x,y)=(-1,0)$ or $\Omega_{\phi}=1$ in the early
universe.  This corresponds to a kinetic energy rather than
matter-dominated universe, a trajectory that disrupts large-scale
structure formation, and hence is not allowed. 

Therefore, we have shown that the blue curve in
Fig.~\ref{fig:sol}\subref{fig:b} is the least constrained viable
trajectory. As argued above, this leads to the constraint $c\lesssim
0.6$.

One might argue that the point $x=y=0$ is repulsive, so that the initial condition for the $\lambda=c=0.6$ trajectory 
is fine-tuned. However, it can be realized without fine-tuning in a model with varying
$\lambda$. As a simple 
 example, consider a potential with two exponential terms 
\begin{align}
  V(\phi)
  &=
  V_1 e^{\lambda_1 \phi / M_{pl}}
  +V_2 e^{\lambda_2 \phi / M_{pl}}
\end{align}
such that  $\lambda \approx \lambda_1 \gg \sqrt{3}$ in
the early universe and switches to $\lambda \approx \lambda_2 =c =0.6$
at some recent point in the past. 
At early times when $\lambda \approx \lambda_1 \gg 1$, $\phi$ rolls
downhill quickly converging to a scaling solution in which
$\Omega_\phi \approx 3/\lambda_1^2 \ll 1$, or $(x,y) \approx
(0,0)$.  At late times when  $\lambda \approx \lambda_2 =c $, the
solution flows to dark energy domination and $(x,y) \rightarrow
(c/\sqrt{6},\sqrt{1- c^2/6})$.  Together, these two stages approximate
the boundary trajectory.  The two exponential model was studied in
detail
in~\cite{Chiba:2012cb}.

The experimental bound we have found for single exponential potential
with constant $\lambda$ agrees reasonably with
the analysis in~\cite{Chiba:2012cb} based on older data. 
A good analytical approximation for the limiting trajectory can be found,
\begin{align}
  x
  &\approx
  \frac{c}{\sqrt{6}}
  \left(
  1-\frac{1-\Omega}{\sqrt{\Omega}}
  \tanh^{-1}(\sqrt{\Omega})
  \right)
  \sim
  \frac23 \frac{c\, \Omega}{\sqrt{6}}
  \label{eq:xapp}
\end{align}
where in the last term 
we have used a first order approximation that will be more convenient. 
This gives a lower bound on $1+w$ for today:
\begin{equation}
  1+w(z=0) \gtrsim
\frac{4}{27} c^2 \Omega_\phi^0 
\label{eq:wbnd}
\end{equation}

The above derivation assumes that the net field excursion in $\phi$ up
until the present is less than $\Delta$, the maximum allowed by
Criterion 1. Indeed for the limiting exponential potential we find
\begin{align}
  \Delta \phi =\sqrt 6 \int x dN \simeq\frac13 c\,{\Omega_\phi^0}
\end{align}
Interestingly, this provides an observational restriction on the
Swampland criterion, namely,
\begin{align}
  \Delta \gtrsim \frac13 c\ \Omega_\phi^0
  \ .
\end{align}

\medskip
\noindent
\section{Future}
The Swampland criteria proposed above also have implications for the
possible futures of our universe given current observational
constraints on the dark energy density and the equation of state. 
The bound derived on the slope of the trajectory in
equation~(\ref{eq:slope}) implies that the value of $x = \dot{\phi}/\sqrt{6}H$ increases in the
immediate future. 

There are three possible future fates for the universe: 

If $\lambda(\phi)$ stays below
$\sqrt{3}$,  $x$ increases approaching the value
$\lambda/\sqrt{6}$. Since the field keeps rolling,  Criterion 1 is violated after a finite time. This would imply a breakdown
of the effective field theory. The universe would enter a new phase in which a large
number of previously massive states become light. This happens when,
\begin{align}
  \Delta \phi =\sqrt{6} \int x dN = \Delta \;
  {\rm or} \; N < \frac{\Delta}{\sqrt{6} x(N=0)} 
  \\ \Rightarrow
  N\lesssim
  \frac{3\Delta}{2c\ \Omega_\phi^0}
\end{align}
where $N$ is the number of e-foldings.   In other words, the new phase would begin within a few Hubble times into the future.

Alternatively, a contrived possibility is that $\lambda(\phi)$ grows
very rapidly in the very near future before $\phi$ rolls
significantly further downhill. In this case, the universe enters a
new phase  in which the field speeds up and $w$ grows, ending the
cosmic acceleration phase ($w < -\frac13$) before $N
\sim\frac{3\Delta}{2c\Omega_\phi^0}$ $e$-folds have passed. 

Finally, we can imagine a situation where the potential reaches zero
or a negative value before $N \sim \frac{3\Delta}{2c  \Omega_\phi^0 }$
$e$-folds have passed.    This clearly marks a different kind of new
phase of the universe in which supersymmetry might be restored or the
universe might enter a phase of contraction.   

Here we have enumerated the possible long-term futures of our universe
given current observational and theoretical constraints. The fact that
in all scenarios the universe survives in its current state at most
for a time period of ${\cal O}(1)$ $e$-fold is a novel explanation for
the observed age of the universe being of order the current Hubble
time.  The bound on
$N$ implies there is a maximal age to the universe as we know it.  The
indicator that the end of the current phase is near is signaled by the
onset of cosmic acceleration as we have already witnessed in our
universe.  Observers cannot exist in a universe where dark energy
dominates for a long time because the universe changes character
first.  A typical observer would measure an age comparable to the
lifetime of the universe today based on the Swampland analysis. 

\medskip
\noindent
\section{The Role of Observations}

The Swampland criteria are based on experience to date in finding
constructions that are consistent with a quantum theory of gravity.
These suggest a maximal field excursion $\Delta \sim {\cal O}(1)$ and
minimum logarithmic slope $|\nabla_{\phi} V|/V> c \sim {\cal O}(1)$.
The ``$\sim$'' in these conditions indicates there is some looseness,
although notably there do not exist rigorously proven examples in hand
where $c$ is as small as 0.6, as required to satisfy current
observational constraints on dark energy.
This is exciting because it means that experiments are already
sensitive enough to put pressure on string theory and the Swampland. 

Based on what we already know observationally about dark energy and
what is shown here, there are clearly important challenges for
theorists:  find rigorous constructions with $c \le 0.6$ that are
consistent with quantum gravity and not in the Swampland
(for example, see~\cite{Cicoli:2012tz,
Gupta:2011yj,
Panda:2010uq,
Kaloper:2003qn,
Hellerman:2001yi,
Choi:1999wv,
Choi:1999xn} for an attempt to embed quintessence in string theory). 
For
inflation, not only the bounds on $\{c,\Delta\}$ are somewhat in
tension
with them being ${\cal O}(1)$
but also almost all current inflationary models studied in the
literature have $|\nabla_\phi V|/V \rightarrow 0$ at one or more
values of $\phi$. 

The situation also provides an opportunity for observers and
experimentalists:  improving  bounds on the dark energy equation of
state, $w(z)$ for $0<z<1$ could push
the limit on $c$ down significantly, further increasing the tension
between observations and Criterion 2.   Similarly, improved
constraints on the tensor-to-scalar ratio $r$ based on CMB
observations will add to the tension between inflationary models and
the Swampland criteria, perhaps pointing to other theories to explain
the large-scale properties of the universe consistent with quantum
gravity.  Finally, we have motivated the possibility of a direct
coupling of the quintessence dark energy field  $\phi$ to dark matter.
Since the quintessence field is rolling today, and perhaps picking up
speed, it is worth searching for evidence that the properties of dark
matter (mass, couplings, etc.) are changing e.g.~by looking for
apparent violations of the equivalence principle in the dark sector.

\section{The Cosmological Constant  Problem and Quintessence}
So far, we have studied what the observational implications of the
Swampland criterion are. It is also worth commenting on how these
criteria change our perspective on the cosmological constant problem
and quintessence.
If there are no de Sitter vacua, then the cosmological constant
problem takes on a different character. 

We can parametrize the scalar
potential for $V>0$ without loss of generality as
\begin{align}
  V(\phi)
  &=
  %V_0 \exp\left(-\frac{1}{M_{pl}}\int_0^\phi \lambda(\phi') d\phi'\right)
  V_0 \exp\left(-\int_0^\phi \frac{d\phi'}{M (\phi')} \right)
\end{align}
such that $|M_{pl} \nabla_\phi V/ V| \equiv M_{pl}/M(\phi)$, where we
have restored explicit factors of $M_{pl}$ for illustration. Then,
assuming that the initial value of the potential is
${\cal{O}}(M_{pl}^4)$, we find
that 
\begin{align}
  \int_0^\phi \frac{d\phi'}{M(\phi')}
  =
  \Delta \phi \left\langle{\frac{1}{M(\phi)}}\right\rangle
  = \log \frac{\Lambda}{M_{pl}^4} 
  \simeq 280
\end{align}
where $\Lambda = (2\ \rm{meV})^4$ is the value of the dark energy today. 
We see
that for field excursions of $\mathcal{O}(M_{pl})$, we need
$\langle M(\phi) \rangle\sim10^{-2}M_{pl}$. Therefore, even though $|M_{pl}\nabla_\phi V/ V|
< 0.6$ today, it should have been larger in the earlier universe. 

Intriguingly, $\langle M(\phi) \rangle \sim 10^{-2}M_{pl}$ points to an interesting scale, the
GUT scale. 
%For example, if the potential is, 
%\begin{align}
%  V(\phi) &= V_0 \exp \left(-\frac{\phi}{M_\phi}\right)
%\end{align}
%Then the scale $M_\phi = M_{pl}/\lambda \sim 10^{16}\ \rm{GeV}$. 
While $M(\phi)$
is a scale in the dark sector, 
it may be related to $M_{GUT}$ naturally as they are both set by the
geometry of the string compactification. 
This then leads to an expectation for dark
energy today of the form
\begin{align}
  \Lambda
  &=
  M_{pl}^4
  \exp\left(-\frac{\# M_{pl}}{M_{GUT}} \right)
\end{align}
with $\#$ denoting the uncertainty in the field excursion up to the
present, expected
to be $\mathcal{O}(1)$ in Planck units. 
Thus, this line of reasoning leads to a relation between the
cosmological constant and the GUT scale. 
%Here the
%hierarchy arises because it is natural to have fixed values 
%for $\nabla_\phi V/V$ 
%leading to an exponential function and hierarchy.

{\bf Acknowledgments}:
We would like to thank R. Daly, A. Ijjas, H. Ooguri, L. Spodyneiko and
C. Stubbs for valuable discussions. 
PA would like to thank the Kavli Institute for Theoretical Physics for their hospitality during the completion of this work. 
The work of PA is supported by
the NSF grants PHY-0855591 and PHY-1216270.  PJS thanks the NYU Center
for Cosmology and Particle Physics and  the Simons Foundation  {\it
  Origins of the Universe Initiative} for support during his leave at
  NYU. PJS is supported by the DOE grant number DEFG02-91ER40671.  The
  work of CV is supported in part by NSF grant PHY-1067976.
\medskip
\noindent

\bibliographystyle{utphys} 
\bibliography{ref}

\end{document}